\documentclass[structabstract]{aa}  
\usepackage{graphicx}
\usepackage{txfonts}
\usepackage{natbib}
\bibpunct{(}{)}{;}{a}{}{,}   

\def\ks{\hbox{$K_{\rm s}$}}
\begin{document}
   \title{Stellar Ages through the Corners of the Boxy Bulge\thanks{Based on data taken at the ESO/VLT Telescope, within the 
   observing programme 081.B-0489(A).}\thanks{The derived photometric catalogs are available in electronic form at the CDS via anonymous ftp to cdsarc.u-strasbg.fr (130.79.128.5) or via http://cdsweb.u-strasbg.fr/cgi-bin/qcat?J/A+A }}


   \author{E. Valenti
          \inst{1}
	  M. Zoccali
	  \inst{2}
 	  A. Renzini
          \inst{3} 
	  T.~M. Brown
	  \inst{4} 
          O. Gonzalez
          \inst{5}
	  D. Minniti
	  \inst{2}
	  Victor P. Debattista
	  \inst{6}
	  L. Mayer
	  \inst{7}
         }

   \institute{European Southern Observatory, 
              Karl Schwarzschild\--Stra\ss e 2, 
	      D\--85748 Garching bei M\"{u}nchen, Germany.\\
              \email{evalenti@eso.org}
	  \and
	      Instituto de Astrof\'{i}sica, Facultad de F\'{i}sica, Pontificia Universidad Cat\'olica de Chile, Av. Vicu\~na Mackenna 4860, Santiago, Chile.
         \and
             INAF\--Osservatorio Astronomico di Padova, Italy
	  \and
	      Space Telescope Science Institute, 3700 San Martin Drive, Baltimore, MD 21218, USA
	  \and
              European Southern Observatory,
              Casilla 19001 Santiago, Chile
          \and
              Jeremiah Horrocks Institute, University of Central Lancashire, Preston PR1 2HE, UK
	  \and
	     Institute for Theoretical Physics, University of Zürich, Winterthurestrasse 190, CH-8057 Zurich, Switzerland 
            }

   \date{}

 
  \abstract
 {} 
   {In some scenarios for the formation of the Milky Way bulge the stellar population at the edges of the boxy bulge may be younger than those on the minor axis, or close to the Galactic center. So far the only bulge region where deep color\--magnitude diagrams have been obtained is indeed along the minor axis. To overcome this limitation, we aim at age\--dating the bulge stellar populations far away from the bulge minor axis.}
   {Color\--magnitude diagrams and luminosity functions have been obtained from deep near\--IR VLT/HAWK\--I images taken at the two Southern corners of the boxy bulge, i.e., near the opposite edges of the Galactic bar. The foreground disk contamination has been statistically removed  using a pure disk field observed with the same instrument and located approximately at similar Galactic latitudes of the two bulge fields, and $\sim 30^\circ$ in longitude away from the Galactic center. 
For each bulge field, mean  reddening and distance are determined using the position of red clump stars, and the metallicity distribution is derived photometrically using the color distribution of stars in the upper red giant branch.}
   {The resulting  metallicity distribution function of both fields peaks around [Fe/H] $\sim -0.1$~dex, with the bulk of the stellar population having a metallicity within the range: $-1$~dex $\lesssim$[Fe/H]$\lesssim +0.4$~dex, quite similar to that of other inner bulge fields.
Like for the inner fields previously explored, the color\--magnitude diagrams of the two bar fields are consistent with their stellar population being older than $\sim 10~Gyr$, with no obvious evidence for the presence of a younger population. 
}
   {The stellar population of the corners of the boxy bulge appears to be coeval with those within the innermost $\sim 4^\circ$ from the Galactic center.}

   \keywords{Galaxy: --
                Bulge --
                formation --
	     Techniques: -- photometric
               }
\authorrunning{Valenti et al.}
   \maketitle
%

\section{Introduction}
The formation of the central regions of disk galaxies that we call galactic bulges remains a debated topic in galaxy evolution. At very early times ($z\ga 4$) the rapid and chaotic congregation of merging gas-rich lumps may have established a first, slowly-rotating nucleus around which disk galaxies could start to grow. Later, at $z\sim 2$, when an extended gas-rich disk had formed, fed by a quasi-steady {\it cold streams}, instabilities are likely to have caused fragmentation into massive star-forming clumps that dynamical friction may have driven to coalesce at the bottom of the potential well, thus adding a rotating component \citep[e.g.][]{noguchi99,Immeli04,carollo07,elmegreen08}. 
At even later times ($z\la 1$), a dynamical instability of the now dominant stellar disk may have caused the inner part of the disk to collapse into a rotating bar \citep[e.g.,][]{shen10}, from now on sharing the same potential well with previously formed structures.

Alternatively, the recent cosmological hydrodynamical Eris simulations show that bar instabilities can start early, at $z > 3$, quickly establishing a compact rotating component with the structural properties of a pseudobulge in low redshift galaxies, which weakens due to mergers and strengthens at low z, after the merging activity has faded away \citep{guedes13}. In this latter scenario it is only the last part of the evolution that resembles the conventional secular bulge formation scenario since the initial growth phase is highly dynamical. These simulations show that, owing to its early assembly, the pseudobulge is dominated by stars with ages of 10 Gyr or more.

The relative role and timing of these processes (and possibly others) in establishing the galactic bulges we observe in the local Universe is yet to be assessed,  and likely may differ from one galaxy to another. Our own Galactic bulge offers a unique opportunity to investigate in detail to which extent these processes have operated in one specific case. Indeed, extensive photometric, astrometric, and spectroscopic data have started to accumulate, allowing us to measure age, chemical compositions, and kinematic properties for a large number of bulge stars, tempting us to identify bulge components that may be ascribed to one of the mentioned processes or another.

Photometric and spectroscopic observations have established that the metallicities of bulge stars span over a broad range, from [Fe/H] $\simeq -1$ to well above solar \citep[e.g.][]{zoccali03,FMR06,zoccali08,johnson11,johnson+13}, with a gradient of $\sim 0.6$ dex/kpc for $|b|>4^\circ$ along the bulge minor axis \citep{zoccali08}. The $\alpha$ elements appear to be enhanced relative to Solar proportions, with [$\alpha$/Fe]$\simeq +0.4$ at low metallicities, then decreasing to reach near-solar reference value at high metallicities \citep{rich07,FMR07,zoccali06,lecureur07,johnson11,gonzalez11a}, following a trend  quite similar to the thick disk stars \citep{melendez08,alvesbrito10,gonzalez11a}.

Star counts and radial velocity studies have shown that the bulge is significantly elongated (it is, in fact, a bar) with axial ratios $\sim 1:0.41:0.38$ and an inclination between $29^\circ$ and $32^\circ$ with respect to the line of sight to the Galactic center \citep[][and references therein]{Cao+13bar}.

\begin{figure}[t]
   \centering
   \includegraphics[width=8cm]{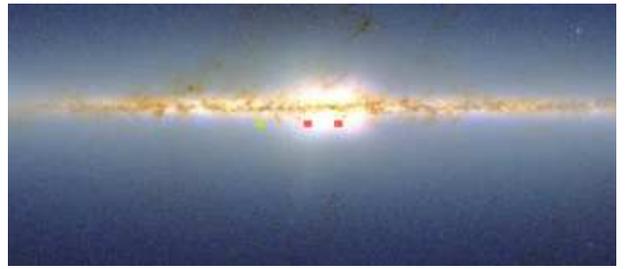}
   \caption{2MASS image of the Galactic bulge \citep{skrutskie06}. The peanut shape of the Bulge is clearly visible.  From left to right (decreasing longitude), the location of the fields observed with HAWK\--I is approximately marked with filled squares, red for the two Bulge fields and green for the disk control field.}
              \label{2massmap}%
    \end{figure}

As shown in Figure~\ref{2massmap}, the bulge isophotes exhibit a boxy structure that appears to be a natural product of the bar formation process. Indeed, simulations show that a substantial vertical thickening occurs preferentially near the ends of the bar \citep[e.g.][]{combes90,pfenniger91,raha91,debattista04}. Moreover, counts of red giant branch clump stars in several directions towards the outer bulge have revealed the presence of an X-shaped structure \citep{manuXshape,nataf10,saito11}, similar to that present in the boxy bulges of other galaxies \citep{bureau06}.

One can argue that the successive operation of such bulge forming processes, via early merging and later instabilities in gas-rich and gas-poor disks, should have left an imprint in the age distribution of bulge stars, and in correlations among their ages, compositions, and kinematical properties. Yet, deep color-magnitude diagrams (CMDs) of bulge fields have so far revealed a uniformly old stellar population, older than $\sim 10$ Gyr, with no appreciable trace of younger stars \citep{ortolani95,kuijken02,zoccali03,sahu06,clarkson08,clarkson11}, thus arguing for an early formation of the bulge, as opposed to a slow, secular growth extended over a large fraction of the Hubble time. Some exceptions are younger stars suggested by microlensing follow\--up spectroscopy \citep{bensby13}.

However, one limitation of these studies is that they all explored small fields (called {\it windows}) that lie on or near the bulge minor axis\footnote{One exception is the study by Brown et al. (2010) where one field well removed from the minor axis was also observed.}, a region that may be avoided by the orbits  of stars added to the bulge by the bar instability. In this respect, the boxy shape being directly linked to the bar formation, the four corners of the boxy bulge appear instead to be the ideal places to search for younger stellar populations.

To explore this possibility, we have obtained deep near-IR images of two fields near the edges of the boxy bulge, located as shown in Figure~\ref{2massmap}. Here we present the resulting CMDs from which we constrain the stellar ages. 

An accurate dating of the bulge component allows one to gauge at which lookback time (i.e., at which redshift) one should  look for possible analogs of the Milky Way, when their bulge formation processes were about to start, well ongoing, or already concluded. Indeed, with an age of $\sim 10$ Gyr or older, it is at $z\ga 2$ that such analogs can be searched, or at lower redshifts if a several Gyr younger components were to be found. Much progress has been achieved in recent years in mapping galaxy populations at
these redshifts, with large, rotating disks being quite common, along with other more compact, velocity dispersion dominated galaxies \citep[e.g][]{genzel+06,FostSch+09,mancini+11}.
In such disks star formation activity often peaks at the center, as judged from the surface brightness distribution of the H$\alpha$ emission, hence suggesting rapid, in situ bulge formation. Moreover, these disk galaxies are characterized by high velocity dispersions ($\ga 50$ km s$^{-1}$), and are far more gas rich than local spirals of the same mass, with gas fraction of $\sim 50\%$ or more \citep{daddi+10,tacconi+10}.  Thus, it is among galaxies of this kind that we may find progenitors of the Milky Way galaxy and its bulge, suggesting that bulge formation took place at a time when the Galaxy was a gas-rich, actively star forming object in which stellar dynamics was only one of the actors at play. Dating stellar populations of the bulge in  various directions would set important constraints on its formation and on the cosmic epochs (redshifts) where to look for analog progenitors. This is indeed the focus of the present paper. 

The paper is organized as follows: the observations and data reduction are presented in Section~\ref{obs}, while in Section~\ref{cmd} we show the derived CMDs and discuss their general properties. Section~\ref{compar} is devoted to discussing the main photometric and physical differences of the two bulge regions. The photometric metallicity distribution of the bulge fields is presented in Section~\ref{mdf}, while in Section~\ref{age} we derive the bulge age. Finally, Section~\ref{comm} is devoted to discuss and summarize our results.

\section{Observation and data reduction}
\label{obs}
Two bulge fields, BUL\--SC29 and BUL\--SC9, located at the opposite edges of the Galactic bar (see Table~\ref{tab-log} and Figure~\ref{2massmap}) were observed with HAWK\--I on Yepun (VLT\--UT4) telescope at ESO Paranal Observatory through the $J$ and $Ks$ filters.  HAWK\--I is a near\--IR imager with a pixel scale of $0.106"$/pix and a total field of view of $7.5'\times7.5'$.  

Both fields have been previously observed by the OGLE\--II survey, hence in this work we adopted their name: BUL\--SC9 and BUL\--SC29 \citep{udalski00, sumi04}.
 BUL\--SC9 is located at the nearest bar edge whereas BUL\--SC29 is at the far end with respect to the Sun position.
The two fields have slightly different latitudes because of the need to avoid very bright sources (i.e. Ks$\la 8$) which cause detector persistence problems (see HAWK\--I User Manual for further details).

Exposures of a disk control field located at $(l,b)=(29.9^{\circ}, -3.9^{\circ})$ were also obtained in order to estimate, and statistically subtract the contamination by disk stars towards the two bulge fields.

Each bulge and disk frame is the combination of 20 exposures each 1.3~sec long. The observation has been repeated with a random dithering pattern until reaching the total exposure time listed in Table~\ref{tab-log}. A random dithering technique was applied with a jitter box width of $30"$ both for background subtraction and to cover the small cross\--shaped gap of $\sim15"$ width between the four detectors. The observations were taken in service mode during Periods 81 and 83 (i.e., in 2008 and 2009) with an average seeing on image of 0.5"$\pm$0.1 (see Table~\ref{tab-log} for more details). Notice from Table~\ref{tab-log} that the planned $J$-band observations for BUL\--SC29 field were not completed, resulting in shallower images compared to the data obtained for the other fields.
 
The processing of the raw data was performed with standard IRAF routines.  For both filters, we obtained a sky image by median combination of the dithered images and subtracted it from each frame.  To normalise the pixel\--to\--pixel response, all frames were then divided by a normalised twilight flat. Finally, all the flat- and sky-corrected frames have been averaged in a single image for each of the two filters. In what follows we will refer to {\it frame} as the combination of each of these sets.

  \begin{figure}[b]
  \centering
  \includegraphics[height=9cm]{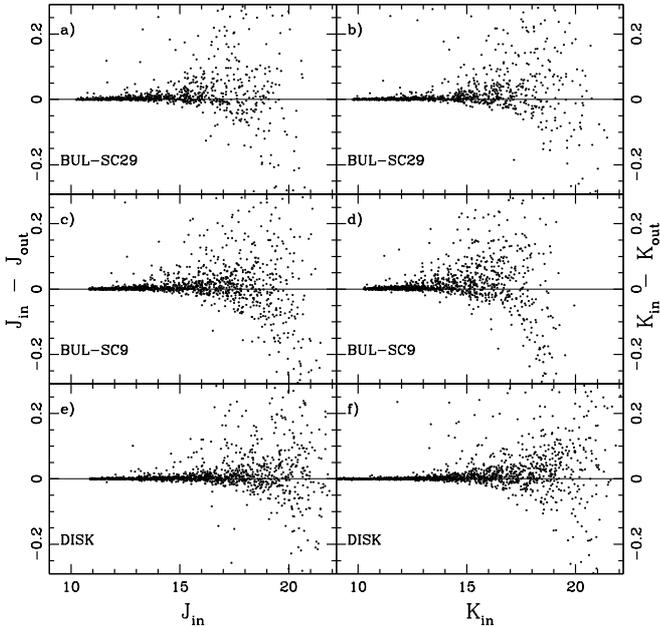}
     \caption{Difference between the input and output magnitude of the artificial star experiments
     on the HAWK\--I frames for the observed bulge and disk fields. Notice the typical asymmetry of these differences, with blending with fainter sources making some stars brighter than they are.}
        \label{compl}
  \end{figure}
%

\begin{table}[ht]
\caption{\label{tab-log}Galactic coordinates, image quality  and Log of the observed fields}
\centering
\begin{tabular}{lccccc}
\hline\hline 
&&&&&\\
Name & l & b & Filter & Exp.Time& FWHM\\
&[deg]&[deg]&& [sec]& [arcsec]\\
\hline
&&&&&\\
BUL\--SC9 & 10.3 & -4.2 & J & 3120 &0.6 \\
&&&$\ks$& 10920 & 0.7\\
&&&&&\\
BUL\--SC29 & -6.8 & -4.7 & J & 1560 & 0.6\\
&&&$\ks$& 10920 & 0.4\\
&&&&&\\
DISK & 29.9 & -3.9 & J & 3120 & 0.5\\
&&&$\ks$& 9360 &0.4\\
&&&&&\\
\hline\hline
&&&&&\\
\end{tabular}
\end{table}
We carried out standard photometry, including PSF modeling, on each frame using DAOPHOTII and ALLSTAR \citep{daophot}. For each observed field, a photometric catalog listing the instrumental $J$ and $\ks$ magnitudes was obtained by cross\--correlating the single\--band catalogs. We used more than 200 stars in common between the derived catalogs and 2MASS data to perform the absolute calibration and astrometrization onto the 2MASS photometric and astrometric system \citep{skrutskie06}.

  Completeness and error estimates were derived via artificial\--star experiments. About 4,000 stars were randomly added to the original bulge and disk frames with magnitudes and colors consistent with the red giant branch (RGB) and main sequence (MS) instrumental fiducial lines. In order to avoid artificially increasing the crowding in each independent experiment the artificial stars were added along the corners of an hexagonal grid \citep[see][for more technical details]{zoccali00}. Figure~\ref{compl} shows the results of the completeness experiments as the difference between the input and output $J$ and $\ks$ magnitudes of the artificial stars for all the observed fields. The asymmetry in the distributions shown in Figure~\ref{compl} is due to occasional blending of two or more stars that the photometric package is not able to resolve. This effect is present over the whole magnitude range, but becomes quite prominent around $J\sim 17$, i.e., at the level of the MS turnoff (TO) of the dominant population, an effect that must be quantitatively taken into account when trying to identify a possible intermediate age population, or set limits to it.  The simulations demonstrated that the photometry is more than 50\% complete above $J\sim$20 and $\ks\sim$19.
%
\section{The Color\--Magnitude Diagrams}
\label{cmd}
\begin{figure}[t]
   \centering
   \includegraphics[height=9.5cm, width=9.5cm]{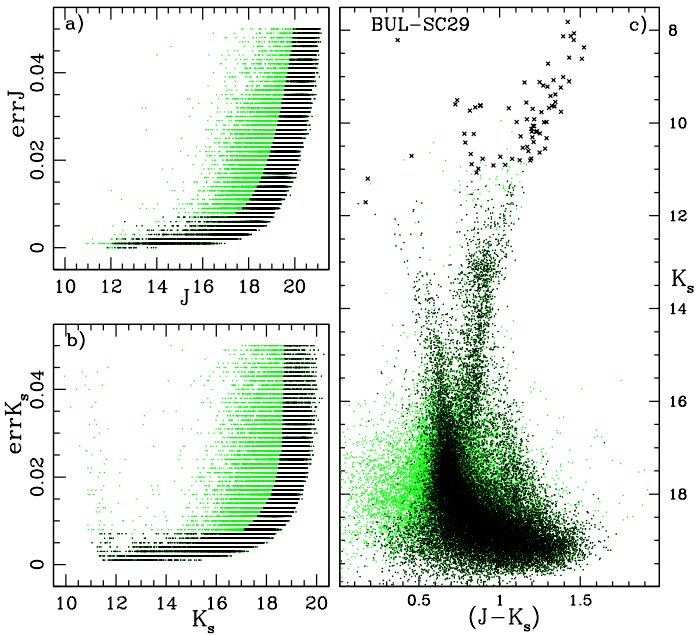}
   \caption{{\it Right panel c) :} the derived ($\ks, J-\ks$) CMD of all stars measured in the BUL\--SC29 field, including both bulge and disk contributions. Black dots refer to stars with errors smaller than the selection applied in panels {\it a} and {\it b} (see text for more details), with green points referring to stars whose errors exceed such limits.
 Black crosses show stars from 2MASS,  which are saturated in the HAWK\--I images.}
              \label{sc29cmd_raw}%
    \end{figure}

\begin{figure}[t]
   \centering
   \includegraphics[height=9.5cm, width=9.5cm]{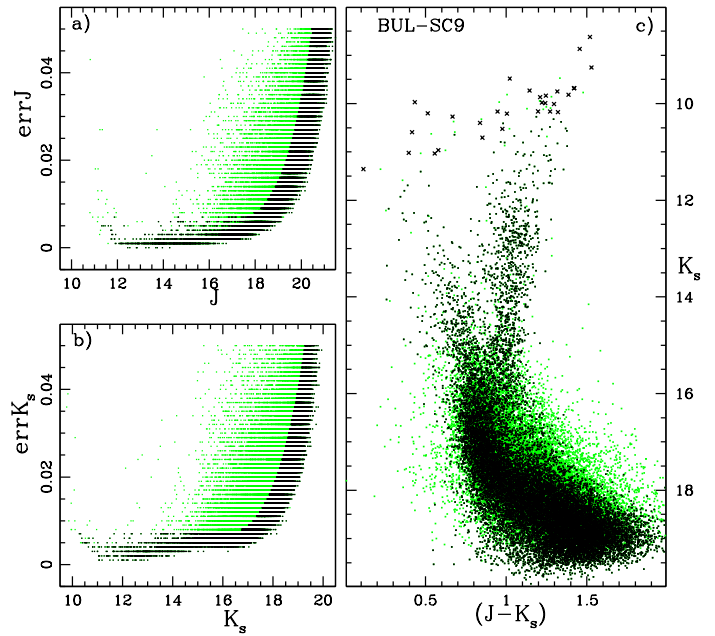}
   \caption{As Figure~\ref{sc29cmd_raw} but for the observed bulge BUL\--SC9 field.}
              \label{sc9cmd_raw}%
    \end{figure}

The derived ($\ks, J-\ks$) CMD of the observed bulge fields, BUL\--SC29 and BUL\--SC9, is shown in panel {\it c)} of Figures~\ref{sc29cmd_raw} and \ref{sc9cmd_raw}, respectively (green dots). 
In both cases, the large number of detected stars saturates the plots for $\ks\ga 16$ making difficult to distinguish the different evolutionary sequences, such as the subgiant branch (SGB) and the MS\--TO. To improve the CMDs we then excluded stars with errors too large J and $\ks$. In doing this, we apply an error cut (see panels {\it a)} and {\it b)}) following the lower envelope of the Poisson error distribution, hence excluding all stars with errors larger than that expected at their respective magnitude levels. The result of this procedure is a cleaner CMD (black dots in panel {\it c)}) in which the color distribution of stars fainter than $\ks$=16 is much narrower. 
Our photometry turns out to be deep enough to properly sample a good fraction of the RGB, from about 1 magnitude above the horizontal branch (HB) down to a couple of magnitudes below the MS\--TO. The saturation magnitude limit occurs around $\ks\sim$11, hence to recover the brightest part of the RGB we have combined our photometry with 2MASS data (cross symbols in panels {\it c)}).

Both fields have very similar CMDs; the major difference is, however, the spread in the various evolutionary sequences. In the case of BUL\--SC9 all sequences appear to be quite broad because of the larger  differential reddening and larger distance dispersion along the line of sight  (see Section~\ref{compar} for more details). As expected, the characteristic double red clump signature of the X\--shape structure is not present in any of the derived CMDs. In fact, as demonstrated by \cite{manuXshape} the X\--shape bulge structure is visible at latitudes $|b|\ge5.5^{\circ}$, and ranging over $\sim13^{\circ}$ in longitude and 20$^{\circ}$ in latitude.
The lower main sequence of  the BUL\--SC29 field is actually broader, due to the poorer S/N in the $J$ band.
The bulge HB clump is clearly visible at $\ks\sim$13 and ($J-\ks$)$\sim$1, as is the RGB which can be easily traced down to $\ks\sim$16.5 as a distinct and separate sequence. Conversely, the SGB is barely detectable being heavily contaminated by the foreground disk stars. The vertical blue sequences departing from the bulge MS\--TO and from the HB clump and extending upwards correspond, respectively, to the foreground disk MS, and its  disk red clump descendants (see Section~\ref{decont} for more details).  

In Figures~\ref{sc29cmd}a and \ref{sc9cmd}a we compare the bulge field CMDs with a 1~Gyr isochrone \citep{pietrinferni06} for a solar metallicity population and helium abundance Y=0.273 by using a distance modulus and reddening correction of $ (m-M)_0$=14.57, 14.12 and $E(B-V)=0.38$, 0.65 for BUL\--SC29 and BUL\--SC9, respectively (see Section~\ref{compar} for details on reddening and distance). As can be clearly noticed, a young ($\sim 1$ Gyr) population belonging to the bulge would have a much bluer color than the disk MS appearing  in these CMDs. 
Increasing the age, the isochrone turnoff color and luminosity get progressively redder and fainter. Therefore to investigate the possible presence of intermediate\--age population in the bulge it is necessary to remove the foreground disk stars from the CMDs. Actually, the MS of disk stars hits the bulge CMD right on top of its MS\--TO, hence hampering a reliable age measurement.


\subsection{Disk Decontamination}
\label{decont}

    \begin{figure}[t]
   \centering
   \includegraphics[height=9.5cm, width=9.5cm]{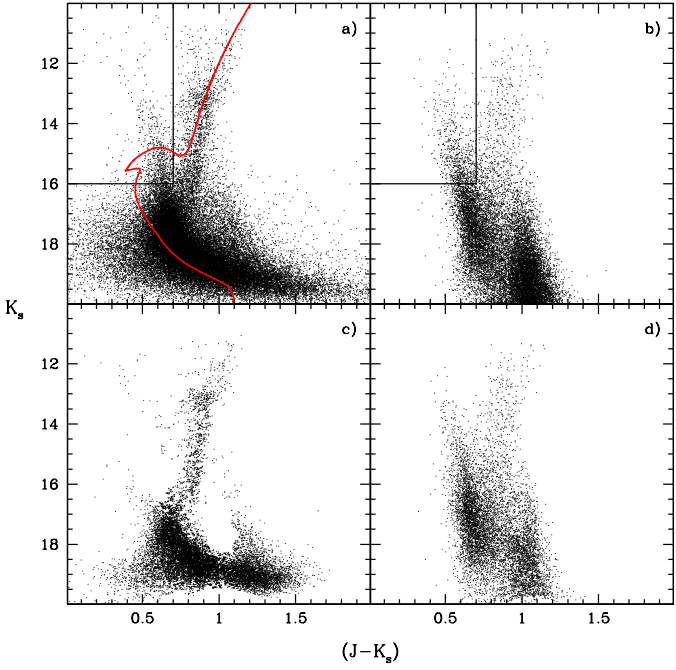}
   \caption{{\it Panel a):} The observed CMD of the bulge BUL-SC29 field . The red solid line is the 1~Gyr isochrone for a solar metallicity population \citep{pietrinferni06}. {\it Panel b):} The observed CMD of the disk control field $\sim 30^\circ$ away from the Galactic center. The region within the box has been used to normalize the number of disk stars seen through the bulge line of sight. {\it Panel c):} The CMD of the bulge BUL-SC29 field as statistically decontaminated from the disk population. {\it Panel d):} The CMD of the stars that were statistically subtracted from the CMD in panel a) in order to obtain the decontaminated CMD shown in panel c). }
              \label{sc29cmd}%
    \end{figure}
    
    \begin{figure}[t]
     \centering
     \includegraphics[height=9.5cm, width=9.5cm]{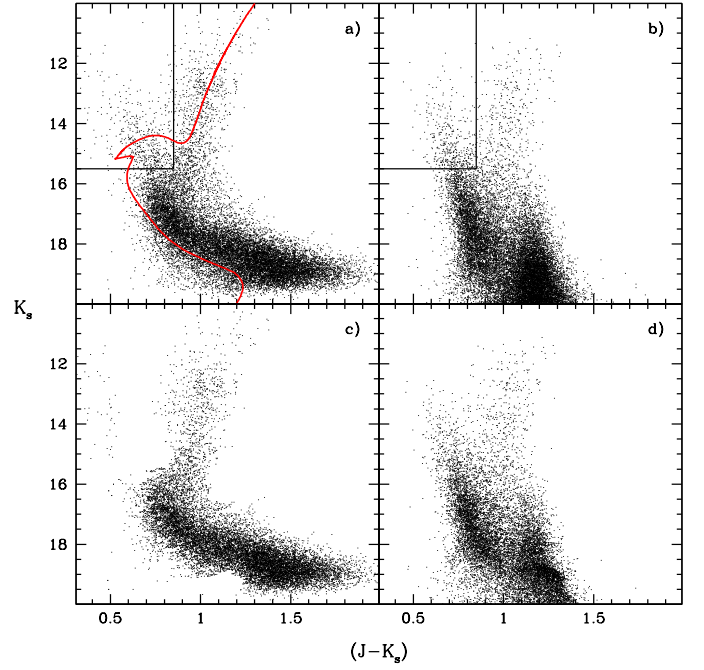}
     \caption{The same as Figure~\ref{sc29cmd} but for the bulge BUL\--SC9 field.}
            \label{sc9cmd}%
    \end{figure}

In order to proceed with age dating of our two bulge fields we must then try to remove as many disk stars as possible from their CMDs. The best decontamination procedure would consist of selecting proper motion members, such as in \cite{kuijken02} and \cite{clarkson11}, even though the result depends on disk kinematics assumptions. As we do not have suitable first epoch data, we adopted a statistical disk decontamination as in \cite{zoccali03}, which has the advantage over the proper motions method to be free from kinematics biases. Moreover, being less time consuming in terms of observations the statistical approach is generally more suitable to decontaminate large areas. The contamination of bulge CMDs from foreground disk stars depends strongly on the latitude of the line of sight and therefore 
 we have used a control disk field located approximately at the same Galactic latitude of the bulge fields but  about 30$^\circ$ away from the Galactic center (see Table~\ref{tab-log}). 
The derived ($\ks, J-\ks$) CMD of the control disk field is shown in Figures~\ref{sc29cmd}b and \ref{sc9cmd}b. This procedure assumes that the disk CMD is the same in the  two bulge fields and in the field at $l\simeq 30^\circ$, which of course may not be strictly the case. To reduce as much as possible the bias that field-to-field differences would induce  we checked for difference in extinction between the fields. In the case of the BUL\--SC29 field, to match the location of the blue, nearly vertical disk sequence, the disk control field was shifted by 0.21 mag in color, and in magnitude by the corresponding $A_{\rm K}=0.35\Delta E(B-V)$ \citep{cardelli+89}. We found no appreciable difference in extinction between BUL\--SC9 and the disk\--control fields, hence no shift in color and magnitude has been applied. For each bulge field, we have  then calculated the bulge\--to\--disk normalization factor which gives us the number of stars to be removed from the bulge CMD for each  given disk star. To do so, after the mentioned reddening adjustment,  we use a region of the CMDs where no bulge stars are present,  i.e., within the box in the upper left corners in Figures~\ref{sc29cmd}a,b and \ref{sc9cmd}a,b. In the disk control field the number of stars within the box is 1.3 and 1.2 times less than in the corresponding box of BUL\--SC29 and BUL\--SC9 field, respectively. Therefore, for each star in the disk CMD of Figures~\ref{sc29cmd}b and \ref{sc9cmd}b we have selected the closest star in the bulge CMDs of Figures~\ref{sc29cmd}a and \ref{sc9cmd}a, and subtracted it adopting these normalization factors and the star completeness found in Section~\ref{obs}. Following the prescription of \citet{zoccali03}, the distance on the CMD from a disk star to each bulge star is computed as $d=\sqrt{[5\times\Delta(J-\ks)]^2+\Delta J^2]}$, where the factor 5 has the function of giving to color differences a weight close to that of magnitude differences, as the color range is much smaller than that spanned by magnitudes. 
For the BUL\--SC29 field, the decontamination procedure worked quite well as demonstrated in Figure~\ref{sc29cmd} panels c) and d), which respectively show the CMD of the disk\--free bulge field and the CMD of the stars statistically removed. 
In the BUL\--SC29 cleaned CMD one can now clearly follow most of the bulge evolutionary sequences: from the HB clump and lower RGB down to the SGB and the MS\--TO. What looks as an excess of stars at the faint end of the {\it cleaned} bulge MS, at $\ks\geq$17 and $(J-\ks)\geq$1.2, is instead an artifact due to the different $J$-band magnitude limit reached in the bulge and disk CMDs, as the broader color distribution of the shallower data (of the bulge field) is not matched by the deeper disk data.
As far as the BUL\--SC9 field is concerned (see panel {\it c)} of Figure~\ref{sc9cmd}), the disk decontamination is more accurate at faint magnitudes being the photometry equally deep in the bulge and in the disk control field. 
However, the presence of larger differential reddening in this field combined with larger distance spread along the line of sight and a lower density make the CMD sequences at and above MS\--TO broader than in the case of the  BUL\--SC29 field, hence the MS\--TO is less sharply defined. In fact, when we use one of the analytic model from \cite[][equation (3e)]{dwek+95} with the parameters recently derived by \cite[][i.e. $x_0=0.68$~Kpc, $y_0=0. 28$~Kpc, $z_0=0.26$~Kpc]{Cao+13bar}, we calculate that the bulge star density in BUL\--SC9 is a factor $\sim 2.2$ smaller than in BUL\--SC29.


\section{The Luminosity Function of the Two Fields}
\label{compar}
\begin{figure}
   \centering
   \includegraphics[height=9.5cm, width=9.5cm]{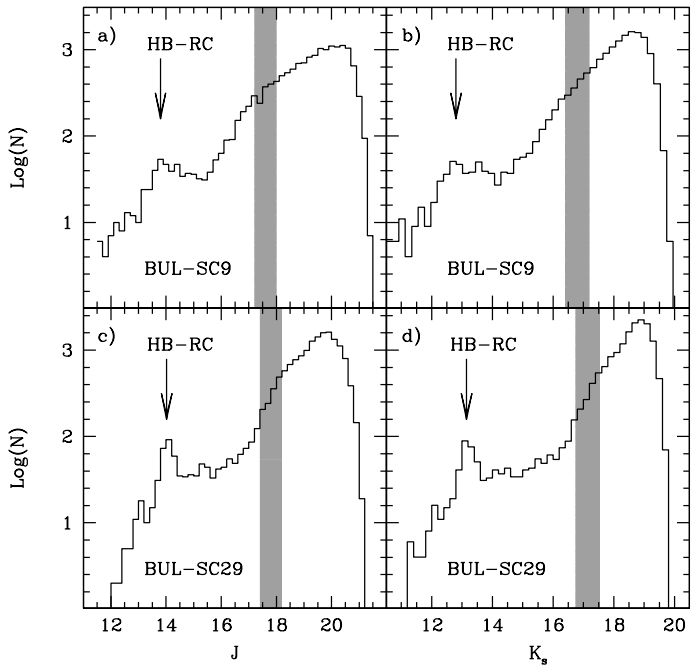}
   \caption{LFs of the observed bulge fields as obtained from the disk\--decontaminated CMDs. The black arrows mark the location of the HB\--RC whereas the grey areas refer to the approximate location of the MS\--TO.}
              \label{lf}%
    \end{figure}

We have used the statistically decontaminated catalogs derived in Section~\ref{decont} to construct the $J$- and $\ks$-band luminosity functions (LF) of the two bulge fields. In case of BUL\--SC29, we also applied a color cut in the low MS to avoid the artifact due to the different J\--band magnitude limit reached in the bulge and disk CMDs (see \S~\ref{decont}). The result is shown in Figure~\ref{lf}, where the location in magnitude of the observed HB red clump (RC) is marked with a black arrow (see also Table~\ref{tab-results}).
In order to quantify the reddening along the two bulge lines of sight we have adopted a differential method based on the comparison between the HB\--RC color, $(J-\ks)^{RC}$, of the observed fields and that of Baade's Window. In fact, the color shift needed to overlap the HB\--RC of two bulge fields with comparable age and metallicity is only a function of the reddening. 
According to \citet{gonzalez11b}, in the Baade's Window  the HB\--RC is located at $\ks^{RC}$=13.15 and $J^{RC}$=14.11. Using the \citet{cardelli+89} extinction coefficients and assuming $E(B-V)=0.55$ for Baade's Window \citep{gonzalez11b}  an average reddening $E(J-\ks)=0.20\pm$0.03 (i.e. $E(B-V)=0.38$) has been derived for the BUL\--SC29 field and $E(J-\ks$)=0.34$\pm$0.05 (i.e. $E(B-V)=0.65$) has been derived for the BUL\--SC9 field. These estimates nicely agree with the corresponding values from \citet{gonzalez12} which are  reported in Table~\ref{tab-results}.

We have derived the mean distance of the two bulge fields by using these reddening estimates and adopting $M_{\rm K}^{\rm RC}=-1.55$ which is the intrinsic $\ks$ magnitude of the HB\--RC for a population 10~Gyr old  of solar metallicity \citep{pietrinferni+04}. This value of $M^{\rm RC}_{\rm K}$was adopted by \cite{gonzalez11b} to map the reddening over the whole bulge and is within 0.02 mag from the empirical calibration of \cite{vanhel07}. The derived mean intrinsic distance moduli of the BUL\--SC29 and BUL\--SC9 fields are listed in Table~\ref{tab-results}. They indicate that BUL\--SC29 field is located at the far edge of the bar and the BUL\--SC9 field is located at the near edge. This is consistent with the larger distance spread in the BUL\--SC9 field indicated by its  broader CMD sequences. Indeed, we notice from Figure~\ref{lf} that the RC feature is much sharper in the BUL\--SC29 field than it is in the BUL\--SC9
field, as expected.

\begin{table*}[ht]
\caption{\label{tab-results} Derived parameters for the observed bulge fields}
\centering
\begin{tabular}{lccccccccc}
\hline\hline 
&&&&&&&&&\\
Field & E(J\--K)$^a$ & E(J\--K)$^b$& E(B\--V)$^c$ & (m\--M)$_0^d$& J$^{RC}$ & $\ks^{RC}$ & J$^{TO}$ & $\ks^{TO}$ & $<[Fe/H]>$\\
&&&&&&&&&\\
\hline
&&&&&&&&&\\
BUL\--SC9 &   0.32 & 0.34 & 0.65 & 14.12 & 13.80 & 12.80 & 17.60 &16.80 & -0.10\\
&&&&&&&&&\\
BUL\--SC29& 0.21 & 0.20 & 0.38 & 14.57 & 14.02 & 13.15 & 17.80 & 17.14 & -0.15\\
&&&&&&&&&\\
\hline\hline
&&&&&&&&&\\
\multicolumn{10}{l}{$^a$ Reddening from Gonzalez et al. (2012);  $^b$ Reddening measured in this work;}\\
\multicolumn{10}{l}{$^c$ Assuming the \citet{cardelli+89} extinction coefficients and the E(J\--K) listed in column [3];}\\
\multicolumn{10}{l}{$^d$ Assuming M$_K^{RC}=-1.55$ (see text for more details).}
\end{tabular}
\end{table*}

\section{The Photometric Metallicity Distributions}
\label{mdf}
In this section we derive the metallicity distribution function (MDF) of the two bulge fields by adopting a procedure  similar to that described by \citet{zoccali03} and used in \citet{gonzalez13}.  
For this purpose we need to use only the bright part of the RGB, which however is saturated in the HAWK\--I data and poorly populated as illustrated in Figures~\ref{sc29cmd_raw} and ~\ref{sc9cmd_raw}.  Therefore, we resort to 2MASS data, using the CMD of 30'$\times$30' regions centered on the two bulge fields, together  with an empirical grid of Galactic cluster RGB ridge lines selected from the sample of \citet{valenti+04a}. Such comparisons must be performed in the absolute CMDs [$M_{\rm K}$, $(J-\ks)_0$], hence for the two fields we use the reddening and distance values derived in Section~\ref{compar}, and listed in Table~\ref{tab-results}. We then derive the photometric metallicity estimate for each star from its color by interpolating within the templates grid. 
    \begin{figure}
   \centering
   \includegraphics[height=9cm]{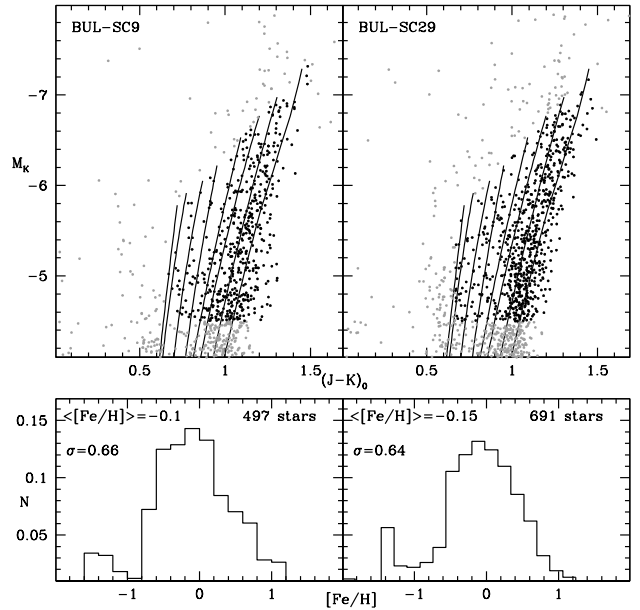}
   \caption{{\it Upper panels:}  The CMDs from 2MASS of the upper RGB for the two $30'\times 30'$ fields including our Hawk-I bulge fields in the absolute [$M_{\rm K}$, ($J-\ks)_0$] plane. The empirical globular cluster RGB templates are over-plotted with metallicities from [Fe/H]$=-$2.1 to +0.35 (see text in Section~\ref{mdf} for details on cluster names and metallicities). Black symbols indicate the stars used to derive the photometric metallicity distribution function. {\it Bottom panels:} Derived photometric metallicity distribution function of the observed bulge fields with a bin size of 0.15~dex.}
   
              \label{fig-mdf}%
    \end{figure}
    

We selected the following globular clusters as empirical RGB templates: M92 ([Fe/H]$=-$2.1), M55 ([Fe/H]$=-$1.61, NGC6752 ([Fe/H]$=-$1.42), NGC362 ([Fe/H]$=-$1.15), M69 ([Fe/H]$=-$0.68), NGC6440 ([Fe/H]$=-$0.50), NGC6528 ([Fe/H]$=-$0.17), and the old open cluster NGC6791 ([Fe/H]$=+$0.35). The selection has been made in order to cover the widest metallicity range in suitable fine steps.

As shown in the upper panels of Figure~\ref{fig-mdf},  the derived MDF of the two bulge fields is obtained considering only RGB stars {\it i)} fainter than the RGB-Tip as derived by \citet{valenti+04b}; {\it ii)} brighter than M$_K = -4.5$, to retain the region of the RGB with the highest sensitivity to metallicity variations, to avoid contamination by red clump stars and to exclude the AGB clump stars, which are predicted to lie around M$_K\sim -2.7$ \citep{pietrinferni+04}; and {\it iii)} excluding stars bluer than ($J-\ks)_0 = 0.7$ to minimize any possible residual contamination by disk stars. 

The derived MDF for the two fields is shown in the lower panels of Figure~\ref{fig-mdf} together with the number of selected stars. The major source of uncertainty in the inferred MDFs is the reddening $E(J-\ks)$ and its spread within each field (which will tend to broaden the distribution). To quantify the errors we also compute the MDF by assuming reddening estimates which differ from those listed in column [3] of Table\ref{tab-results} by $\pm$0.03 and $\pm$0.05 for BUL\--SC29 and BUL\--SC9, respectively. The result is a shift of the peak distribution by $\lesssim$0.15~dex. For further details on the error estimates the reader can refer to \citet{gonzalez13} who derive the photometric metallicity map of the bulge by using the same empirical grid and 2MASS photometric catalog. It should be mentioned that given the uncertainties associated to reddening, depth effects and possible mixing of populations, the adopted method provides a reliable description of the overall metallicity range spanned by the bulk of the population (i.e. mean and dispersion of the distribution). Here we are interested in deriving the metallicity range to be adopted in \S~\ref{age} by the theoretical isochrones to derive the stellar ages.

Within the errors, the derived MDFs are very similar in the two fields. They are quite broad, with the bulk of the stars lying within the metallicity range $-0.8\lesssim$[Fe/H]$\lesssim$+0.6 and with a peak around [Fe/H]$\sim-0.1$. As expected, the peak of the derived MDFs nicely agree with the results by \cite{gonzalez13} who found [Fe/H]=-0.20~dex and -0.15~dex, for BUL\--SC29 and BUL\--SC9 respectively.

Given the limitations of the method, these MDFs are broadly consistent  with the spectroscopic MDFs of the bulge \citep{zoccali08}, and in the case of the  BUL\--SC29 field with the multiband photometric MDF of \cite{brown+10}. In both cases, the secondary peak at low metallicity  is most likely due to a contamination from foreground disk stars in the HB-RC. We conclude that the two bulge fields have indistinguishable MDFs.

\section{Stellar Ages}    
\label{age}
Using the statistically decontaminated CMDs presented in Section~\ref{decont} we now estimate the age of the bulge stellar populations at the two opposite edges of the bar. 

\begin{figure}
   \centering
   \includegraphics[height=9.0cm]{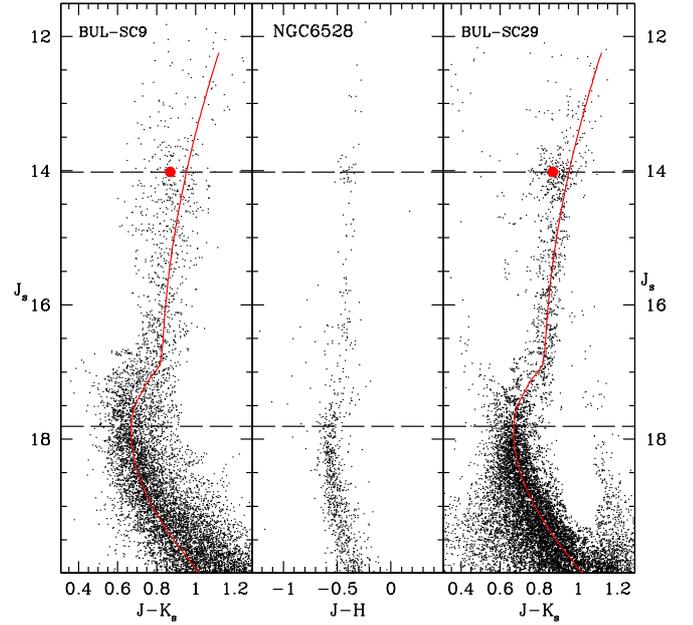}
   \caption{Comparison between the CMDs of the bulge fields observed in this work (left and right panels). The magnitude difference between the HB\--RC and MS\--TO of  the BUL\--SC29 field (dashed horizontal lines) is compared with the same quantity for the BUL\--SC9 field. The CMD of the BUL\--SC9 field has been shifted in magnitude and color to match the location of BUL\--SC29 HB\--RC.  In both these panels the solid red line is the empirical ridge line of the CMD  of  the BUL\--SC29 field. The central panel shows the $J-(J-H)$ CMD of the bulge globular cluster NGC 6528, showing that the luminosity difference between the HB--RC and the MS--TO is virtually identical to that of the two bulge fields.}
   
              \label{confr_age}
    \end{figure}

For the sake of comparison, in Figure~\ref{confr_age} the CMD of BUL\--SC9 has been shifted in magnitude and color to match the location of the HB\--RC in the BUL\--SC29 field, having identified the magnitude of the HB--RCs from Figure~\ref{lf}. The difference between the magnitude of the HB\--RC and the MS\--TO is essentially the same in the two fields as emphasized by the horizontal dashed lines, implying that the stellar population in these two bulge regions are coeval. 
Moreover, to further show the similarity between the two fields in Figure~\ref{confr_age} we have over-plotted in both panels the CMD ridge line for the BUL\--SC29 field (red solid line). As one can easily appreciate, the fiducial ridge line of BUL\--SC29 nicely fits also all the observed CMD sequences of the BUL\--SC9 field, from the MS--TO up to the RGB, thus ruling out any appreciable differences between the two fields in terms of mean metallicity and age. 

The central panel in Figure~\ref{confr_age} shows the $J, (J-H)$ CMD of the bulge globular cluster NGC~6528 from HST/NICMOS data \citep{ortolani01} tranformed into 2MASS photometric system by applying the \citet{brandner+01} color transformations. The $J$-band magnitude has been shifted in order to have the HB-RC of the cluster at the same level as that of the two bulge fields. The metallicity of NGC~6528, as measured from high\--resolution optical and near\--IR spectroscopy, ranges $-0.1\leq$[Fe/H]$\leq+0.07$ \citep{carretta01,zoccali04,origlia05}, i.e., it is very close to the mean metallicity of the two bulge fields.
The figure shows that the $J$-band luminosity difference between the HB-RC and the MS--TO $\Delta J^{\rm HB}_{\rm TO}$ of the cluster is virtually the same as that of the two bulge fields, which ensures that the cluster and the bulk stellar population in the bar fields are virtually coeval. This extends to the edges of the bar the same age dating procedure pioneered by \cite{ortolani95} for the Baade's Window field in the bulge. 
Still, the turnoff region of the BUL\--SC29 field \-- and even more so of the BUL\--SC9 field \-- appears to be more extended in luminosity compared to that of the cluster. There are indeed important differences between cluster and fields, namely, the cluster stars are chemically homogeneous, all at the same distance and suffer relatively little differential reddening being sampled in the small $19''.2\times 19''.2$ NICMOS field of view. Conversely, stars in the two bulge fields span a wide range in metallicity and distance, and are more affected by differential reddening over the much larger $7'.5\times 7'.5$ field of view of HAWK\--I.
Of course, we cannot completely exclude the presence of a minority of metal-rich, few Gyr younger stars (see below).
 \begin{figure*}
   \centering
   \includegraphics[height=14cm]{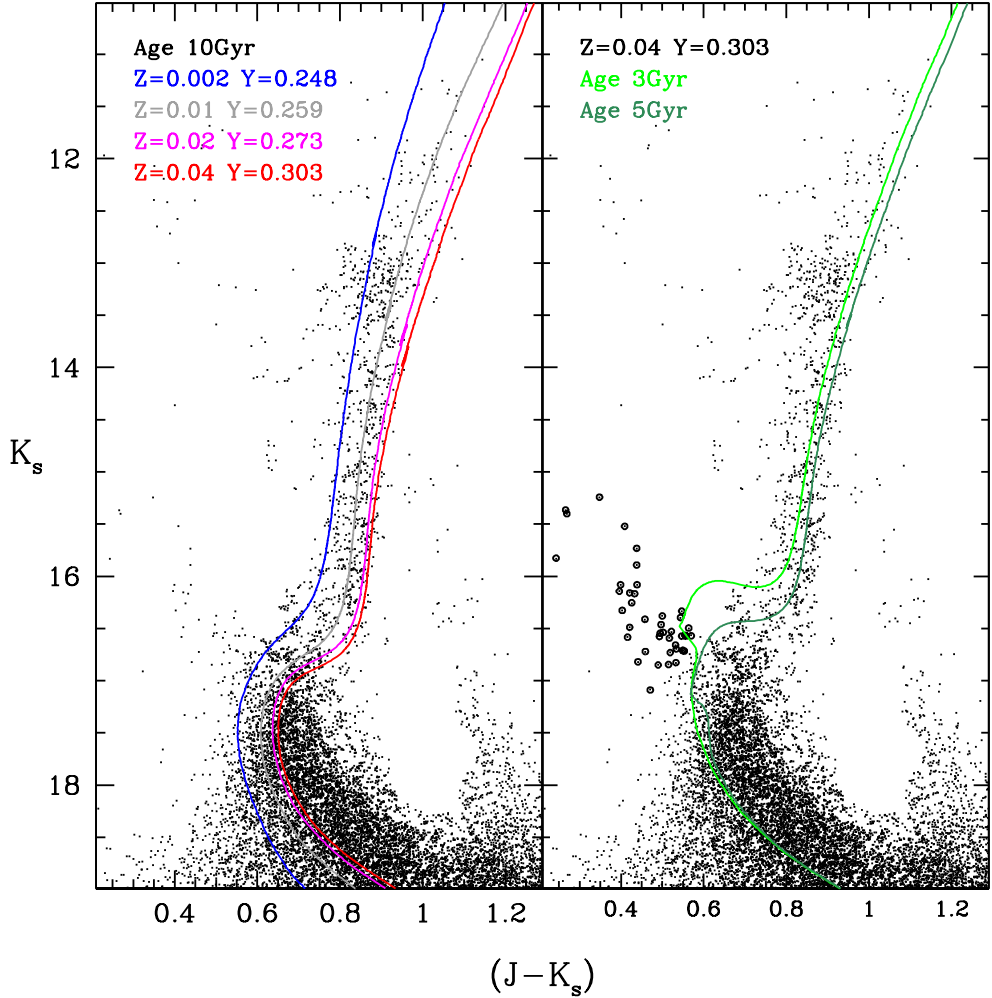}
   \caption{The  CMD of the BUL\--SC29 field with overplotted  theoretical isochrones \citep{valcarce12}. Isochrone ages, metallicities and helium abundances are indicated near the top-left corner of each panel. Sub\--solar isochrones are $\alpha$\--enhanced, solar\--scaled otherwise. The  metallicity range covered by the bulk of the stars in the BUL\--SC29 field is 0.002$\lesssim Z\lesssim$0.060,  as derived in Section\ref{mdf}. Open circles mark stars which are most likely blue stragglers (see text for more details).}
   
              \label{abs_age}%
    \end{figure*}
    
To more accurately quantify this conclusion we consider in more detail the case of the BUL\--SC29 field, which is less affected by differential reddening and distance dispersion effects. The left panel of Figure~\ref{abs_age}  shows that the bulk of the bulge population has an age $\gtrsim$10~Gyr. In fact, when comparing the BUL\--SC29 CMD with 10~Gyr theoretical isochrones of various metallicities,  the observed spread in color is fully consistent with the metallicity dispersion measured in Section~\ref{mdf}. Indeed, most stars have metallicity in the range between [Fe/H]$\sim -1$ (blue solid line) and [Fe/H]$\sim +$0.6 (red solid line),  with a peak around [Fe/H]$\sim -$0.10~dex (grey solid line). Most stars in the BUL\--SC29  field are in fact confined within the two extreme isochrones. Moreover, as we pointed out in Section~\ref{obs}, our artificial stars experiments have shown that the accuracy of our photometry suffers from blending effects. In particular in the MS\--TO luminosity range (i.e. $\ks\sim 17.15\pm0.2$), the blending could affect the star measurements up to $\sim$15\%, leading to a systematic overestimate of the luminosity (see also Figure~\ref{compl}). This effect rapidly increases with decreasing luminosity along the MS, which accounts for the increasing mismatch between the isochrones and the data seen in Figure~\ref{abs_age}. Hence, the observed luminosity dispersion affecting the MS\--TO region can be well accounted when considering the blending effect together with $\sim$0.26~mag 1\--$\sigma$ dispersion due to the distance distribution along the line of sight \citep[see Section 5 in][]{gonzalez13}, with very little room left for the presence of an intermediate age population. 
It is worth emphasizing that blending can make stars brighter but never bluer than the true turnoff. Therefore, the blue cutoff of the MS can put a stronger constraint on age than the measured MS--TO luminosity, providing that an appropriate set of isochrones is used.

This is further illustrated by the comparison between the observed CMD and 3 and 5~Gyr isochrones for super\--solar metallicity, shown in the right panel of Figure~\ref{abs_age}, which demonstrates that there is indeed little room left for the presence of young/intermediate-age populations. 

The spur of stars bluer than $J-\ks\sim 0.6$ and brighter than $\ks\sim 17$, marked with open circles in the right panel of Fig.~\ref{abs_age}, are most likely the blue straggler progeny of binary star evolution, as demonstrated for an inner bulge field by \cite{clarkson11}. In order to verify it we calculated the blue straggler frequency ($S_{BSS}=N_{BSS}/N_{HB}$) as the number of blue straggler ($N_{BSS}$) normalized to the total number of HB stars ($N_{HB}$) used as reference population. $N_{HB}$ has been estimated by using the LFs shown in Fig.~\ref{lf}: by subtracting from the RC\--HB peak the contribution of the RGB stars, previously derived by fitting the LF in the range $J \ge (J^{RC}+1)$ and $J \le (J^{RC}-1) $. We found $S_{BSS}$=0.4, which is consistent with the result of \cite[][i.e. $0.31\le S_{BSS} \le 1.23$]{clarkson11}. Finally, it is worth to mention that our estimate of $S_{BSS}$ should be regarded as a lower limit because the derived  $N_{HB}$ refers only to the number of red HB stars, being the small contribution from blue HB stars very difficult to estimate from our photometry.

As we have mentioned earlier, the BUL\--SC29 field was also observed with HST/WFC3 by \cite{brown+10} who noted that its $J-(J-H)$ CMD is virtually identical to that of three other inner bulge fields, just a few degrees from the Galactic center. Though a detailed age-dating of such fields was beyond the scope of the \citet{brown+10} paper, they note that all such fields are dominated by an old ($\sim 10$ Gyr) population.

\section{Discussion and conclusion}
\label{comm}

The near-IR CMDs we have constructed for the two Southern corners of the boxy bulge indicate, as far as we can say, a uniformly old stellar population ($\sim 10$ Gyr, or older), similar to that of the more central fields close to the minor axis of the bulge, when the same CMD method is employed \citep{ortolani95, zoccali03, brown+10, clarkson11}. The CMDs of these bulge fields are indeed very similar to that of the bulge globular cluster NGC~6528 which has a metallicity very close to the mean metallicity of the bulge fields. In particular, the luminosity difference between the HB\--RC and the MS\--TO is virtually identical as shown in Figure~\ref{confr_age}, with $\Delta J^{\rm HB}_{\rm TO}\simeq 3.8\pm 0.2$, which ensures cluster and fields having essentially the same age within better than $\sim 20\%$. Thus, rather than trying here a detailed isochrone fitting dating of the bulge fields, we report prior age estimates of the bulge cluster NGC~6528.

This is in very good agreement with  results from the recent Eris cosmological simulations \citep{guedes13}, which coincidentally produce a galaxy that by $z=0$ resembles very closely the internal structure and mass distribution of an Sb-Sbc spiral galaxy such as our own Milky Way \citep{guedes11}. This prompts to compare more profound aspects of pseudobulge formation between simulations and detailed observations, such as metallicity evolution and metallicity-age relation, that can in turn aid the interpretation  of current and future datasets. 
Detailed multi-wavelength mock photometry and spectroscopy of the simulated galaxies  will be an important tool in order to make progress in this direction.

Restricting to age estimates  based on HST data, NGC~6528 was given an age 13$\pm 3$ Gyr by \cite{ortolani01} based on NICMOS data, 11$\pm 2$ Gyr by \cite{feltzing02} based on two epoch WFPC2 data (hence using only proper-motion cluster members), 12.6~Gyr by \citet{momany03} based on proper motion members and a combination of WFPC2 and near\--IR data, and finally 12.5 Gyr by \cite{brown+05} based on ACS data.
We conclude that the bulk stellar population of the galactic bar edges (identified with the two corners of the boxy bulge indicated in Figure~\ref{2massmap}) is over $\sim 10$ Gyr old, and such age is indistinguishable from that reported for more inner bulge fields, a few degrees from the Galactic center or lying along the bulge minor axis.  On this basis, we do not find evidence for the age distribution in these fields far off the minor axis being any different from that on the minor axis. Thus, if the bar instability had the effect of injecting disk stars into the bulge, it might have done so a long time ago.

Partly at variance with our present results and all previous ones based on the CMDs of specific bulge fields
\citep{ortolani95, zoccali03, clarkson08, clarkson11, brown+10} is the age-dating of bulge stars based on the
microlensing dwarf project of \citet[][and references therein]{bensby13}. Rather that from magnitudes and colors, their ages are derived from the effective temperature and gravity from the high\--resolution spectra obtained during the microlensing events, then using isochrones in the ($T_{\rm eff}, {\rm log}\, g$) plane. They find that 31 of their 58 microlensed dwarfs ($\sim 53\%$) would be younger than 9 Gyr, with 13 of them ($\sim 23\%$) being younger than 5 Gyr. Whereas the dwarfs more metal\--poor that [Fe/H]$\simeq -0.4$ appear to be uniformly old (older than $\sim 10$ Gyr), the more metal\--rich dwarfs appear to span a very wide range of ages from virtually the age of the Universe down to less than $\sim 2$ Gyr.

There appears to be a discrepancy between the ages inferred from CMD of bulge fields and those derived for the microlensed dwarfs and subgiants. Admittedly, age errors for individual  microlensing events are quite large, and when such uncertainties are taken into account only 3 out of 58 stars  ($\sim 5\%$) within $1\sigma$ are younger than 5 Gyr \citep{bensby13}. This is not too far from the upper limit of $\sim 3\%$ estimated by \cite{clarkson11}.  Of course, each of the two methods has its own advantages and disadvantages. The traditional CMD method deals with very large numbers of stars, and therefore can in principle also reveal trace of young populations. However, the metallicities of individual stars are not known, and one does not know if e.g., some of the stars above the MS\--TO of the $Z=0.060$ isochrone in Figure~\ref{abs_age} are old and of lower metallicity, or they are metal\--rich stars younger than 10 Gyr.  On the other hand, besides forcefully dealing with very small number statistics, the microlensing approach is more heavily dependent on model atmospheres which may introduce systematic effects, especially at high metallicity. However, it has the advantage that the metallicity of each individual star is known.

The Galactic bulge HST/WFC3 Treasury Project \citep{brown+09, brown+10} is designed to overcome the limitations of the CMD method by measuring metallicities photometrically, using a purposely tailored multiband dataset. It will also do so for a very large number of stars around the MS\--TO in four different bulge fields, and using proper motion to select pure bulge stars. For these reasons we believe that a more sophisticated, thorough isochrone dating procedure will be best attempted using the Treasury Project data, rather than the present dataset, where individual metallicities are not known and the disk decontamination of the bulge population is necessarily statistical.

Still, our more solid result is that no appreciable age differences appear to exist between the central regions of the bulge (the innermost $\sim 4^\circ$ from the Galactic center) and the two explored fields at two corners of the boxy bulge, at the edges of the galactic bar.

\begin{acknowledgements}
This material is based upon work supported in part by the National Science Foundation under Grant No. 1066293 and the hospitality of the Aspen Center for Physics.

M.Z. acknowledge support by Proyecto Fondecyt Regular 1110393, the BASAL  Center for  Astrophysics  and  Associated Technologies  PFB-06, the FONDAP  Center for  Astrophysics 15010003, Proyecto Anillo ACT-86 and by  the Chilean Ministry for the Economy, Development,  and  Tourism's  Programa  Iniciativa Cient\'{i}fica  Milenio through grant P07-021-F, awarded to The Milky Way Millennium Nucleus.

A.R. acknowledges the kind hospitality and support of the Astronomy Department of the Pontificia Universitad Catolica when this paper was finally set up.

V.P.D is supported in part by STFC Consolidated grant \# ST/J001341/1.

This work has made use of BaSTI and PGPUC isochrones web tools.
 
This publication makes use of data products from the Two Micron All Sky Survey (2MASS), which is a joint project of the University of Massachusetts and the Infrared Processing and Analysis Center/California Institute of Technology, founded by the National Aeronautics and Space Administration and the National Science Foundation.
\end{acknowledgements}

\bibliographystyle{aa}
\bibliography{mybiblio}

\begin{thebibliography}{65}
\expandafter\ifx\csname natexlab\endcsname\relax\def\natexlab#1{#1}\fi

\bibitem[{{Alves-Brito} {et~al.}(2010){Alves-Brito}, {Mel{\'e}ndez}, {Asplund},
  {Ram{\'{\i}}rez}, \& {Yong}}]{alvesbrito10}
{Alves-Brito}, A., {Mel{\'e}ndez}, J., {Asplund}, M., {Ram{\'{\i}}rez}, I., \&
  {Yong}, D. 2010, \aap, 513, A35+

\bibitem[{{Bensby} {et~al.}(2013){Bensby}, {Yee}, {Feltzing}, {Johnson},
  {Gould}, {Cohen}, {Asplund}, {Mel{\'e}ndez}, {Lucatello}, {Han}, {Thompson},
  {Gal-Yam}, {Udalski}, {Bennett}, {Bond}, {Kohei}, {Sumi}, {Suzuki}, {Suzuki},
  {Takino}, {Tristram}, {Yamai}, \& {Yonehara}}]{bensby13}
{Bensby}, T., {Yee}, J.~C., {Feltzing}, S., {et~al.} 2013, \aap, 549, A147

\bibitem[{{Brandner} {et~al.}(2001){Brandner}, {Grebel}, {Barb{\'a}},
  {Walborn}, \& {Moneti}}]{brandner+01}
{Brandner}, W., {Grebel}, E.~K., {Barb{\'a}}, R.~H., {Walborn}, N.~R., \&
  {Moneti}, A. 2001, \aj, 122, 858

\bibitem[{{Brown} {et~al.}(2005){Brown}, {Ferguson}, {Smith}, {Guhathakurta},
  {Kimble}, {Sweigart}, {Renzini}, {Rich}, \& {VandenBerg}}]{brown+05}
{Brown}, T.~M., {Ferguson}, H.~C., {Smith}, E., {et~al.} 2005, \aj, 130, 1693

\bibitem[{{Brown} {et~al.}(2010){Brown}, {Sahu}, {Anderson}, {Tumlinson},
  {Valenti}, {Smith}, {Jeffery}, {Renzini}, {Zoccali}, {Ferguson},
  {VandenBerg}, {Bond}, {Casertano}, {Valenti}, {Minniti}, {Livio}, \&
  {Panagia}}]{brown+10}
{Brown}, T.~M., {Sahu}, K., {Anderson}, J., {et~al.} 2010, \apjl, 725, L19

\bibitem[{{Brown} {et~al.}(2009){Brown}, {Sahu}, {Zoccali}, {Renzini},
  {Ferguson}, {Anderson}, {Smith}, {Bond}, {Minniti}, {Valenti}, {Casertano},
  {Livio}, {Panagia}, {Vanden Berg}, \& {Valenti}}]{brown+09}
{Brown}, T.~M., {Sahu}, K., {Zoccali}, M., {et~al.} 2009, \aj, 137, 3172

\bibitem[{{Bureau} {et~al.}(2006){Bureau}, {Aronica}, {Athanassoula},
  {Dettmar}, {Bosma}, \& {Freeman}}]{bureau06}
{Bureau}, M., {Aronica}, G., {Athanassoula}, E., {et~al.} 2006, \mnras, 370,
  753

\bibitem[{{Cao} {et~al.}(2013){Cao}, {Mao}, {Nataf}, {Rattenbury}, \&
  {Gould}}]{Cao+13bar}
{Cao}, L., {Mao}, S., {Nataf}, D., {Rattenbury}, N.~J., \& {Gould}, A. 2013,
  ArXiv e-prints

\bibitem[{Cardelli {et~al.}(1989)Cardelli, Clayton, \& Mathis}]{cardelli+89}
Cardelli, J.~A., Clayton, G.~C., \& Mathis, J. 1989, ApJ, 345, 245

\bibitem[{{Carollo} {et~al.}(2007){Carollo}, {Scarlata}, {Stiavelli}, {Wyse},
  \& {Mayer}}]{carollo07}
{Carollo}, C.~M., {Scarlata}, C., {Stiavelli}, M., {Wyse}, R.~F.~G., \&
  {Mayer}, L. 2007, \apj, 658, 960

\bibitem[{{Carretta} {et~al.}(2001){Carretta}, {Cohen}, {Gratton}, \&
  {Behr}}]{carretta01}
{Carretta}, E., {Cohen}, J.~G., {Gratton}, R.~G., \& {Behr}, B.~B. 2001, \aj,
  122, 1469

\bibitem[{{Clarkson} {et~al.}(2008){Clarkson}, {Sahu}, {Anderson}, {Smith},
  {Brown}, {Rich}, {Casertano}, {Bond}, {Livio}, {Minniti}, {Panagia},
  {Renzini}, {Valenti}, \& {Zoccali}}]{clarkson08}
{Clarkson}, W., {Sahu}, K., {Anderson}, J., {et~al.} 2008, \apj, 684, 1110

\bibitem[{{Clarkson} {et~al.}(2011){Clarkson}, {Sahu}, {Anderson}, {Rich},
  {Smith}, {Brown}, {Bond}, {Livio}, {Minniti}, {Renzini}, \&
  {Zoccali}}]{clarkson11}
{Clarkson}, W.~I., {Sahu}, K.~C., {Anderson}, J., {et~al.} 2011, \apj, 735, 37

\bibitem[{{Combes} {et~al.}(1990){Combes}, {Debbasch}, {Friedli}, \&
  {Pfenniger}}]{combes90}
{Combes}, F., {Debbasch}, F., {Friedli}, D., \& {Pfenniger}, D. 1990, \aap,
  233, 82

\bibitem[{{Daddi} {et~al.}(2010){Daddi}, {Bournaud}, {Walter}, {Dannerbauer},
  {Carilli}, {Dickinson}, {Elbaz}, {Morrison}, {Riechers}, {Onodera}, {Salmi},
  {Krips}, \& {Stern}}]{daddi+10}
{Daddi}, E., {Bournaud}, F., {Walter}, F., {et~al.} 2010, \apj, 713, 686

\bibitem[{{Debattista} {et~al.}(2004){Debattista}, {Carollo}, {Mayer}, \&
  {Moore}}]{debattista04}
{Debattista}, V.~P., {Carollo}, C.~M., {Mayer}, L., \& {Moore}, B. 2004, \apjl,
  604, L93

\bibitem[{{Dwek} {et~al.}(1995){Dwek}, {Arendt}, {Hauser}, {Kelsall}, {Lisse},
  {Moseley}, {Silverberg}, {Sodroski}, \& {Weiland}}]{dwek+95}
{Dwek}, E., {Arendt}, R.~G., {Hauser}, M.~G., {et~al.} 1995, \apj, 445, 716

\bibitem[{{Elmegreen} {et~al.}(2008){Elmegreen}, {Bournaud}, \&
  {Elmegreen}}]{elmegreen08}
{Elmegreen}, B.~G., {Bournaud}, F., \& {Elmegreen}, D.~M. 2008, \apj, 688, 67

\bibitem[{{Feltzing} \& {Johnson}(2002)}]{feltzing02}
{Feltzing}, S. \& {Johnson}, R.~A. 2002, \aap, 385, 67

\bibitem[{{F{\"o}rster Schreiber} {et~al.}(2009){F{\"o}rster Schreiber},
  {Genzel}, {Bouch{\'e}}, {Cresci}, {Davies}, {Buschkamp}, {Shapiro},
  {Tacconi}, {Hicks}, {Genel}, {Shapley}, {Erb}, {Steidel}, {Lutz},
  {Eisenhauer}, {Gillessen}, {Sternberg}, {Renzini}, {Cimatti}, {Daddi},
  {Kurk}, {Lilly}, {Kong}, {Lehnert}, {Nesvadba}, {Verma}, {McCracken},
  {Arimoto}, {Mignoli}, \& {Onodera}}]{FostSch+09}
{F{\"o}rster Schreiber}, N.~M., {Genzel}, R., {Bouch{\'e}}, N., {et~al.} 2009,
  \apj, 706, 1364

\bibitem[{{Fulbright} {et~al.}(2006){Fulbright}, {McWilliam}, \&
  {Rich}}]{FMR06}
{Fulbright}, J.~P., {McWilliam}, A., \& {Rich}, R.~M. 2006, \apj, 636, 821

\bibitem[{{Fulbright} {et~al.}(2007){Fulbright}, {McWilliam}, \&
  {Rich}}]{FMR07}
{Fulbright}, J.~P., {McWilliam}, A., \& {Rich}, R.~M. 2007, \apj, 661, 1152

\bibitem[{{Genzel} {et~al.}(2006){Genzel}, {Tacconi}, {Eisenhauer},
  {F{\"o}rster Schreiber}, {Cimatti}, {Daddi}, {Bouch{\'e}}, {Davies},
  {Lehnert}, {Lutz}, {Nesvadba}, {Verma}, {Abuter}, {Shapiro}, {Sternberg},
  {Renzini}, {Kong}, {Arimoto}, \& {Mignoli}}]{genzel+06}
{Genzel}, R., {Tacconi}, L.~J., {Eisenhauer}, F., {et~al.} 2006, \nat, 442, 786

\bibitem[{{Gonzalez} {et~al.}(2011{\natexlab{a}}){Gonzalez}, {Rejkuba},
  {Zoccali}, {Hill}, {Battaglia}, {Babusiaux}, {Minniti}, {Barbuy},
  {Alves-Brito}, {Renzini}, {Gomez}, \& {Ortolani}}]{gonzalez11a}
{Gonzalez}, O.~A., {Rejkuba}, M., {Zoccali}, M., {et~al.} 2011{\natexlab{a}},
  \aap, 530, A54+

\bibitem[{{Gonzalez} {et~al.}(2013){Gonzalez}, {Rejkuba}, {Zoccali}, {Valent},
  {Minniti}, \& {Tobar}}]{gonzalez13}
{Gonzalez}, O.~A., {Rejkuba}, M., {Zoccali}, M., {et~al.} 2013, \aap, 552, A110

\bibitem[{{Gonzalez} {et~al.}(2011{\natexlab{b}}){Gonzalez}, {Rejkuba},
  {Zoccali}, {Valenti}, \& {Minniti}}]{gonzalez11b}
{Gonzalez}, O.~A., {Rejkuba}, M., {Zoccali}, M., {Valenti}, E., \& {Minniti},
  D. 2011{\natexlab{b}}, \aap, 534, A3

\bibitem[{{Gonzalez} {et~al.}(2012){Gonzalez}, {Rejkuba}, {Zoccali}, {Valenti},
  {Minniti}, {Schultheis}, {Tobar}, \& {Chen}}]{gonzalez12}
{Gonzalez}, O.~A., {Rejkuba}, M., {Zoccali}, M., {et~al.} 2012, \aap, 543, A13

\bibitem[{{Guedes} {et~al.}(2011){Guedes}, {Callegari}, {Madau}, \&
  {Mayer}}]{guedes11}
{Guedes}, J., {Callegari}, S., {Madau}, P., \& {Mayer}, L. 2011, \apj, 742, 76

\bibitem[{{Guedes} {et~al.}(2012){Guedes}, {Mayer}, {Carollo}, \&
  {Madau}}]{guedes13}
{Guedes}, J., {Mayer}, L., {Carollo}, M., \& {Madau}, P. 2012, ArXiv e-prints

\bibitem[{{Immeli} {et~al.}(2004){Immeli}, {Samland}, {Gerhard}, \&
  {Westera}}]{Immeli04}
{Immeli}, A., {Samland}, M., {Gerhard}, O., \& {Westera}, P. 2004, \aap, 413,
  547

\bibitem[{{Johnson} {et~al.}(2011){Johnson}, {Rich}, {Fulbright}, {Valenti}, \&
  {McWilliam}}]{johnson11}
{Johnson}, C.~I., {Rich}, R.~M., {Fulbright}, J.~P., {Valenti}, E., \&
  {McWilliam}, A. 2011, \apj, 732, 108

\bibitem[{{Johnson} {et~al.}(2013){Johnson}, {Rich}, {Kobayashi}, {Kunder},
  {Pilachowski}, {Koch}, \& {de Propris}}]{johnson+13}
{Johnson}, C.~I., {Rich}, R.~M., {Kobayashi}, C., {et~al.} 2013, \apj, 765, 157

\bibitem[{{Kuijken} \& {Rich}(2002)}]{kuijken02}
{Kuijken}, K. \& {Rich}, R.~M. 2002, \aj, 124, 2054

\bibitem[{{Lecureur} {et~al.}(2007){Lecureur}, {Hill}, {Zoccali}, {Barbuy},
  {G{\'o}mez}, {Minniti}, {Ortolani}, \& {Renzini}}]{lecureur07}
{Lecureur}, A., {Hill}, V., {Zoccali}, M., {et~al.} 2007, \aap, 465, 799

\bibitem[{{Mancini} {et~al.}(2011){Mancini}, {F{\"o}rster Schreiber},
  {Renzini}, {Cresci}, {Hicks}, {Peng}, {Vergani}, {Lilly}, {Carollo},
  {Pozzetti}, {Zamorani}, {Daddi}, {Genzel}, {Maraston}, {McCracken},
  {Tacconi}, {Bouch{\'e}}, {Davies}, {Oesch}, {Shapiro}, {Mainieri}, {Lutz},
  {Mignoli}, \& {Sternberg}}]{mancini+11}
{Mancini}, C., {F{\"o}rster Schreiber}, N.~M., {Renzini}, A., {et~al.} 2011,
  \apj, 743, 86

\bibitem[{{McWilliam} \& {Zoccali}(2010)}]{manuXshape}
{McWilliam}, A. \& {Zoccali}, M. 2010, \apj, 724, 1491

\bibitem[{{Mel{\'e}ndez} {et~al.}(2008){Mel{\'e}ndez}, {Asplund},
  {Alves-Brito}, {Cunha}, {Barbuy}, {Bessell}, {Chiappini}, {Freeman},
  {Ram{\'{\i}}rez}, {Smith}, \& {Yong}}]{melendez08}
{Mel{\'e}ndez}, J., {Asplund}, M., {Alves-Brito}, A., {et~al.} 2008, \aap, 484,
  L21

\bibitem[{{Momany} {et~al.}(2003){Momany}, {Ortolani}, {Held}, {Barbuy},
  {Bica}, {Renzini}, {Bedin}, {Rich}, \& {Marconi}}]{momany03}
{Momany}, Y., {Ortolani}, S., {Held}, E.~V., {et~al.} 2003, \aap, 402, 607

\bibitem[{{Nataf} {et~al.}(2010){Nataf}, {Udalski}, {Gould}, {Fouqu{\'e}}, \&
  {Stanek}}]{nataf10}
{Nataf}, D.~M., {Udalski}, A., {Gould}, A., {Fouqu{\'e}}, P., \& {Stanek},
  K.~Z. 2010, \apjl, 721, L28

\bibitem[{{Noguchi}(1999)}]{noguchi99}
{Noguchi}, M. 1999, \apj, 514, 77

\bibitem[{{Origlia} {et~al.}(2005){Origlia}, {Valenti}, \& {Rich}}]{origlia05}
{Origlia}, L., {Valenti}, E., \& {Rich}, R.~M. 2005, \mnras, 356, 1276

\bibitem[{{Ortolani} {et~al.}(2001){Ortolani}, {Barbuy}, {Bica}, {Renzini},
  {Zoccali}, {Rich}, \& {Cassisi}}]{ortolani01}
{Ortolani}, S., {Barbuy}, B., {Bica}, E., {et~al.} 2001, \aap, 376, 878

\bibitem[{{Ortolani} {et~al.}(1995){Ortolani}, {Renzini}, {Gilmozzi},
  {Marconi}, {Barbuy}, {Bica}, \& {Rich}}]{ortolani95}
{Ortolani}, S., {Renzini}, A., {Gilmozzi}, R., {et~al.} 1995, \nat, 377, 701

\bibitem[{{Pfenniger} \& {Friedli}(1991)}]{pfenniger91}
{Pfenniger}, D. \& {Friedli}, D. 1991, \aap, 252, 75

\bibitem[{{Pietrinferni} {et~al.}(2004){Pietrinferni}, {Cassisi}, {Salaris}, \&
  {Castelli}}]{pietrinferni+04}
{Pietrinferni}, A., {Cassisi}, S., {Salaris}, M., \& {Castelli}, F. 2004, \apj,
  612, 168

\bibitem[{{Pietrinferni} {et~al.}(2006){Pietrinferni}, {Cassisi}, {Salaris}, \&
  {Castelli}}]{pietrinferni06}
{Pietrinferni}, A., {Cassisi}, S., {Salaris}, M., \& {Castelli}, F. 2006, \apj,
  642, 797

\bibitem[{{Raha} {et~al.}(1991){Raha}, {Sellwood}, {James}, \& {Kahn}}]{raha91}
{Raha}, N., {Sellwood}, J.~A., {James}, R.~A., \& {Kahn}, F.~D. 1991, \nat,
  352, 411

\bibitem[{{Rich} {et~al.}(2007){Rich}, {Origlia}, \& {Valenti}}]{rich07}
{Rich}, R.~M., {Origlia}, L., \& {Valenti}, E. 2007, \apjl, 665, L119

\bibitem[{{Sahu} {et~al.}(2006){Sahu}, {Casertano}, {Bond}, {Valenti}, {Ed
  Smith}, {Minniti}, {Zoccali}, {Livio}, {Panagia}, {Piskunov}, {Brown},
  {Brown}, {Renzini}, {Rich}, {Clarkson}, \& {Lubow}}]{sahu06}
{Sahu}, K.~C., {Casertano}, S., {Bond}, H.~E., {et~al.} 2006, \nat, 443, 534

\bibitem[{{Saito} {et~al.}(2011){Saito}, {Zoccali}, {McWilliam}, {Minniti},
  {Gonzalez}, \& {Hill}}]{saito11}
{Saito}, R.~K., {Zoccali}, M., {McWilliam}, A., {et~al.} 2011, \aj, 142, 76

\bibitem[{{Shen} {et~al.}(2010){Shen}, {Rich}, {Kormendy}, {Howard}, {De
  Propris}, \& {Kunder}}]{shen10}
{Shen}, J., {Rich}, R.~M., {Kormendy}, J., {et~al.} 2010, \apjl, 720, L72

\bibitem[{{Skrutskie} {et~al.}(2006){Skrutskie}, {Cutri}, {Stiening},
  {Weinberg}, {Schneider}, {Carpenter}, {Beichman}, {Capps}, {Chester},
  {Elias}, {Huchra}, {Liebert}, {Lonsdale}, {Monet}, {Price}, {Seitzer},
  {Jarrett}, {Kirkpatrick}, {Gizis}, {Howard}, {Evans}, {Fowler}, {Fullmer},
  {Hurt}, {Light}, {Kopan}, {Marsh}, {McCallon}, {Tam}, {Van Dyk}, \&
  {Wheelock}}]{skrutskie06}
{Skrutskie}, M.~F., {Cutri}, R.~M., {Stiening}, R., {et~al.} 2006, \aj, 131,
  1163

\bibitem[{Stetson(1987)}]{daophot}
Stetson, P.~B. 1987, PASP, 99, 191

\bibitem[{{Sumi}(2004)}]{sumi04}
{Sumi}, T. 2004, \mnras, 349, 193

\bibitem[{{Tacconi} {et~al.}(2010){Tacconi}, {Genzel}, {Neri}, {Cox}, {Cooper},
  {Shapiro}, {Bolatto}, {Bouch{\'e}}, {Bournaud}, {Burkert}, {Combes},
  {Comerford}, {Davis}, {Schreiber}, {Garcia-Burillo}, {Gracia-Carpio}, {Lutz},
  {Naab}, {Omont}, {Shapley}, {Sternberg}, \& {Weiner}}]{tacconi+10}
{Tacconi}, L.~J., {Genzel}, R., {Neri}, R., {et~al.} 2010, \nat, 463, 781

\bibitem[{{Udalski} {et~al.}(2000){Udalski}, {Zebrun}, {Szymanski}, {Kubiak},
  {Pietrzynski}, {Soszynski}, \& {Wozniak}}]{udalski00}
{Udalski}, A., {Zebrun}, K., {Szymanski}, M., {et~al.} 2000, \actaa, 50, 1

\bibitem[{{Valcarce} {et~al.}(2012){Valcarce}, {Catelan}, \&
  {Sweigart}}]{valcarce12}
{Valcarce}, A.~A.~R., {Catelan}, M., \& {Sweigart}, A.~V. 2012, \aap, 547, A5

\bibitem[{Valenti {et~al.}(2004{\natexlab{a}})Valenti, Ferraro, \&
  Origilia}]{valenti+04a}
Valenti, E., Ferraro, F.~R., \& Origilia, L. 2004{\natexlab{a}}, MNRAS, 351,
  1204

\bibitem[{Valenti {et~al.}(2004{\natexlab{b}})Valenti, Ferraro, \&
  Origilia}]{valenti+04b}
Valenti, E., Ferraro, F.~R., \& Origilia, L. 2004{\natexlab{b}}, MNRAS, 354,
  815

\bibitem[{{van Helshoecht} \& {Groenewegen}(2007)}]{vanhel07}
{van Helshoecht}, V. \& {Groenewegen}, M.~A.~T. 2007, \aap, 463, 559

\bibitem[{{Zoccali} {et~al.}(2004){Zoccali}, {Barbuy}, {Hill}, {Ortolani},
  {Renzini}, {Bica}, {Momany}, {Pasquini}, {Minniti}, \& {Rich}}]{zoccali04}
{Zoccali}, M., {Barbuy}, B., {Hill}, V., {et~al.} 2004, \aap, 423, 507

\bibitem[{{Zoccali} {et~al.}(2000){Zoccali}, {Cassisi}, {Frogel}, {Gould},
  {Ortolani}, {Renzini}, {Rich}, \& {Stephens}}]{zoccali00}
{Zoccali}, M., {Cassisi}, S., {Frogel}, J.~A., {et~al.} 2000, \apj, 530, 418

\bibitem[{{Zoccali} {et~al.}(2008){Zoccali}, {Hill}, {Lecureur}, {Barbuy},
  {Renzini}, {Minniti}, {G{\'o}mez}, \& {Ortolani}}]{zoccali08}
{Zoccali}, M., {Hill}, V., {Lecureur}, A., {et~al.} 2008, \aap, 486, 177

\bibitem[{{Zoccali} {et~al.}(2006){Zoccali}, {Lecureur}, {Barbuy}, {Hill},
  {Renzini}, {Minniti}, {Momany}, {G{\'o}mez}, \& {Ortolani}}]{zoccali06}
{Zoccali}, M., {Lecureur}, A., {Barbuy}, B., {et~al.} 2006, \aap, 457, L1

\bibitem[{Zoccali {et~al.}(2003)Zoccali, Renzini, Ortolani, Greggio, Saviane,
  Cassisi, Rejkuba, Barbuy, Rich, \& Bica}]{zoccali03}
Zoccali, M., Renzini, A., Ortolani, S., {et~al.} 2003, A\&A, 399, 931

\end{thebibliography}

\end{document}